\documentclass[twocolumn,10pt,superscriptaddress]{revtex4-1}
\usepackage{mathtools}
\usepackage{amsfonts}
\usepackage[all]{xy}
\usepackage{tikz}
\usetikzlibrary{arrows, decorations.markings}
\usepackage{subfig}
\usepackage{cmbright}
\usepackage{hyperref}
\usepackage{afterpage}

\begin{document}

\title{Knoto-ID: a tool to study the entanglement of open protein chains using the concept of knotoids}

\author{Julien Dorier} 
\affiliation{Vital-IT, SIB Swiss Institute of Bioinformatics, 1015 Lausanne, Switzerland}
\author{Dimos Goundaroulis}
\affiliation{Center for Integrative Genomics, University of Lausanne, 1015 Lausanne, Switzerland}
\affiliation{Swiss Institute of Bioinformatics, 1015 Lausanne, Switzerland}
\author{Fabrizio Benedetti}
\affiliation{Center for Integrative Genomics, University of Lausanne, 1015 Lausanne, Switzerland}
\affiliation{Vital-IT, SIB Swiss Institute of Bioinformatics, 1015 Lausanne, Switzerland}
\author{Andrzej Stasiak}
\affiliation{Center for Integrative Genomics, University of Lausanne, 1015 Lausanne, Switzerland}
\affiliation{Swiss Institute of Bioinformatics, 1015 Lausanne, Switzerland}
  
  \begin{abstract}  
 \noindent {\footnotesize The authors wish it to be known that, in their opinion, the first two authors should be regarded as joint First Authors.} \\
  
  \noindent \textbf{Abstract:}  The backbone of most proteins forms an open curve.
    To study their entanglement, a common strategy consists in searching for the presence of knots in their backbones using topological invariants.
    However, this approach requires to close the curve into a loop, which alters the geometry of curve.
    Knoto-ID allows evaluating the entanglement of open curves without the need to close them, using the recent concept of knotoids which is a generalization of the classical knot theory to open curves.
    Knoto-ID can analyse the global topology of the full chain as well as the local topology by exhaustively studying all subchains or only determining the knotted core.\\
    \textbf{Availability and Implementation:}  Knoto-ID is written in C++ and includes R (\href{www.R-project.org}{www.R-project.org}) scripts to generate plots of projections maps, fingerprint matrices and disk matrices.
    Knoto-ID is distributed under the GNU General Public License (GPL), version 2 or any later version and is available at \href{https://github.com/sib-swiss/Knoto-ID}{https://github.com/sib-swiss/Knoto-ID}. A binary distribution for Mac OS X and Linux with detailed user guide and examples can be obtained from \href{https://www.vital-it.ch/software/Knoto-ID}{http://www.vital-it.ch/software/Knoto-ID}.\\
  \textbf{Contact:} \href{julien.dorier@sib.swiss}{julien.dorier@sib.swiss}

\end{abstract}

\maketitle

\begin{figure*}[!tbp]
\centerline{\includegraphics{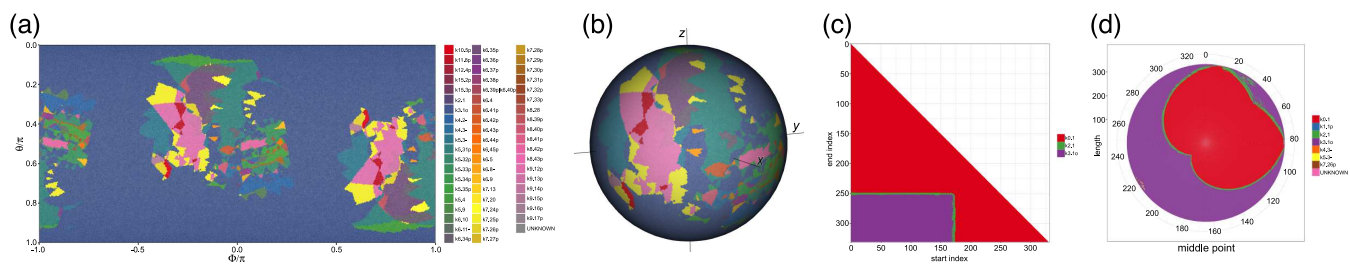}}
\caption{ (a) Projection map on a plane using spherical coordinates or (b) on a globe using Cartesian coordinates. (c) Fingerprint matrix and (d) a disc matrix. This figure was generated with Knoto-ID using the protein 3KZN (\cite{shi2006}) as input.}\label{fig:01}
\end{figure*}

\section{Introduction}

The observation that protein backbones can form knots(\cite{mans}) initiated numerous studies of their nature and potential advantages or disadvantages that they may provide (e.g. \cite{virnau, dabrstas}). In this context, it was important to classify protein knots in terms of their topology. 

A knot is a closed curve in 3-dimensional space that doesn't intersect itself and it can be freely deformed as long as it does not pass through itself (\cite{adams}). However, the backbone of many biomolecules and specifically of many proteins correspond to open spatial curves and so, in strict topological sense, such curves are classified as unknotted. Until recently, the only way to study the topology of an open protein chain was to first close them and then proceed with the study of the entanglement. Of course, closing the chain alters its  geometry. In 2012, V. Turaev introduced the concept of knotoids as a generalization of the classical knot theory to open knots (\cite{turaev}). Knotoids were studied further by N. G\"ug\"umc\"u and L. H. Kauffman (\cite{guka}).  As a consequence, a number of studies emerged that implemented this new mathematical tool in the analysis of a protein backbone (\cite{dennis, dgound, dgound2}).

 In this note we introduce Knoto-ID, a command line tool that is able to analyse and classify  open spatial curves using this new mathematical concept. Moreover, we provide the possibility of closing an open 3D curve, if a knot analysis is required, with either a direct closure (e.g. \cite{taylor}) or using the uniform closure technique (e.g. \cite{sulk}). This note focuses on individual open protein chains, however Knoto-ID can be used to analyse any open linear conformation in 3-space such as chromosomes (\cite{siebert2017}), synthetic polymers, random walks.

\section{Implementation}
To analyse a protein, the coordinates of the $C\alpha$ atoms of the protein backbone have to be extracted from a .pdb file downloaded from the PDB (\href{www.rcsb.org}{www.rcsb.org}, \cite{pdb}) and then stored to a text file, which is the input of Knoto-ID. The backbone is then placed inside a large enough enclosing sphere. Each point of the sphere defines a direction of projection. For a given direction of projection, two infinite lines are introduced that pass from each of the termini of the curve and are parallel to the chosen direction. A triangle elimination method based on the KMT algorithm (\cite{koniaris, taylor}) is then applied to simplify the curve while preserving the its underlying topology with respect to the two parallel lines.
A knotoid diagram is obtained by projecting the curve on an oriented surface (a plane or a sphere).
Finally, a topological invariant is evaluated on the knotoid diagram. For curves projected on a sphere the topological invariant is the Jones polynomial for knotoids (\cite{turaev}), while for curves projected on a plane it is the Turaev loop bracket polynomial (\cite{turaev}) (see the Knoto-ID user guide for a brief description of the theory). The knotoid type corresponding to the resulting polynomial can be optionally obtained using a list knotoid types distributed with Knoto-ID.
Different choices of projection directions may yield different  diagrams and so one has to sample an adequate number of projection directions in order to approximate the spectrum of knotoid types that corresponds to the spatial curve. The spectrum of knotoid types can be visualized using projection maps generated by Knoto-ID (Figure~\ref{fig:01}a and \ref{fig:01}b).
Knoto-ID is also able to handle closed chains as input, or to create a closed chain from an open one using either direct and uniform closure.

Knoto-ID is also able to analyse all subchains of a given curve to produce a fingerprint matrix (\cite{yeates}), for the case of open chains or a disc matrix (\cite{rawdon}), for the case of closed chain (see Figure~\ref{fig:01}c and \ref{fig:01}d). In addition, Knoto-ID can also find the knotted core of the chain, which is the shortest subchain obtained by progressively altering the length of the input chain by 1 point without changing the dominant knot or knotoid type in the process.

\section{Conclusion}
Knoto-ID is the first tool that is able to handle, analyse as well as classify open linear conformations in 3-space such as proteins in terms of their topology  without requiring them to be closed into a loop, using the concept of knotoids.\vspace*{-17pt}

\section*{Acknowledgements}
This is the Author's Original Version of the  article has been accepted for publication in Bioinformatics Published by Oxford University Press. 

We thank Louis H. Kauffman for fruitful discussions, Frédéric Sch\"utz for his advice on Knoto-ID packaging. We also thank Eric Rawdon, Elizabeth Annoni and Nicole Lopez for kindly providing the list of projections distributed with Knoto-ID.\vspace*{-2pt}

\section*{Funding}
The work was funded in part by Leverhulme Trust (RP2013-K-017) and by the Swiss National Science Foundation (31003A-138267), both credited to Andrzej Stasiak. 

\end{document}